# Digital competency of educators in the virtual learning environment: a structural equation modeling analysis

**S M Hizam[1], H Akter[2], I Sentosa[3] and W Ahmed[4]**

[1,2,3,4] UniKL Business School (UBIS), Universiti Kuala Lumpur, Kuala Lumpur, Malaysia

Email: sheikhmhizam@unikl.edu.my[1]

**Abstract.** This study integrates the educators' digital competency (DC), as an individual characteristic construct of the task-technology fit (TTF) theory, to examine a better fit between Moodle using and teaching task, and to investigate its effect on both Moodle's utilization and their task performance. For assessing our proposed hypotheses, an online survey was conducted with 238 teaching staff from different departments of universities in Malaysia. Using Structural Equation Modelling (SEM), our analysis revealed that all the proposed components (i.e., technology literacy, knowledge deepening, presentation skills, and professional skills) of digital competency significantly influenced the TTF. The Task-Technology Fit was also found as an influential construct, which positively and significantly affected both Moodle's utilization and teachers' task performance. Besides, Moodle's utilization was confirmed to be a substantial determinant of the performance impact. In the end, this study included limitations and future directions based on how the study's contribution can support academics and practitioners for assessing and understanding what particular components of digital competency impact TTF, which in turn may influence the system's utilization and performance impact.

## 1. Introduction

It is said that merely change is the constant for a sustainable life. The pre-emptive strategies to cope up with such disruptions and transition challenges are deemed timely necessitate [1]. The global pandemic circumstances have compelled the non-digital mindset to become the technological inclined in professional life in almost every sector. The education segment is the most effective one with a complete shift of the learning process from physical classrooms and labs to a virtual environment. This unexpected transition is succeeded by only those (i.e., teachers, students, and institutions) who were ready to embrace the change by digital skills enhancement. The decision to engage in the learning and development process to comply with ongoing variation generally relies on an individual's behavior [2]. Organizational training programs are ineffective without comprehending the individual's digital literacy and boosting the competency in digital tools and techniques usability and its usufructs. Usually, all the educators are not trained and aimed for the virtual learning environment due to the less requirement and learning subject scenarios but the pandemic impact on educational discourse has let the force adoption of Virtual Learning Environment (VLE) across the academia from school level to universities. In this situation, the role of the individual digital characteristic of educators (i.e., teacher, lecturer, etc.) is a matter of significance in conforming to the education system sustainability. Through VLE, educators engross learning and teaching facilities towards the learners. As academic institutions provide the VLE platform to "teach from home", the educator's behavior towards VLE tools entails for their productive performance [3]. However, there is limited research on considering the individual's digital competence fit towards performing the academician tasks in a technological environment during the pandemic contingency phase. Therefore, this study aims to build the task technology fit on educator's digital competency towards the use of VLE application. To assess the academician's performance behavior by technology fit scenario, a causal study based on a theoretical foundation can better conclude the inferences.

One of the broadly validated theories within academia is the task-technology fit (TTF) theory, a well-established theory of behavioral factors that precisely explains individual characteristics of using technology [4]–[6]. According to Omotayo and Haliru [7], individuals' characteristics portray an individual's attributes that influence their interactions for using information systems (IS) technology. The authors further counted that such attributes cover demographic aspects, attitudes, personality traits, and technology proficiencies (digital competencies). Studies showed that a better understanding of users' characteristics regarding a good fit of information systems might support determining their systems' usage and carrying out their work roles more efficiently and inherently [6], [8], [9]. Within this context, constructs deliberated as individual characteristics in our research are teachers' digital competencies, including technology literacy, knowledge deepening, presentation skills, and professional skills of using VLE (Moodle), resulting in a good fit for Moodle on their task completion. A few researchers have endeavored to study using individuals' skills and abilities as a precursor for TTF [6], [10]. Hence, it is needed to look at broader and more diverse horizons to bring more understanding of the user capabilities [6]. Based on the TTF model, the literature revealed that digital competency plays an important role in understanding the ability of the end-user to use any information system (IS) and how the system is fit to accomplish required tasks, which, in turn, impact the task performance [11]. However, limited researches have applied the digital competency notion in VLE (Moodle) usage among educators in Malaysian higher education institutions' perspectives under the tenet of TTF theory.

With innovation in Information and Communications Technology (ICT) systems impacting teaching and learning strategy, higher education institutions are concerned about learning management systems (LMS) to provide online courses frequently. Moodle, the more popular open-source of LMS, is pretty intuitive of effective teaching-learning processes [12]. To enfold the digital competency for the educators, this study underlined the digital competency from UNESCO's defined particulars. The "ICT Competency Framework for Educators" by UNESCO includes digital literacy (technology literacy), knowledge deepening, didactic instructions (presentation skills), and professional skills, which teachers need while using any ICT tools [13]. In the current global context of educational transformation, higher education institutes in Malaysia also impacted by virtual learning[14]. Under this premise, our study goes yonder to understand and investigate the teachers' digital competency towards using Moodle and its effect on their task-performance in the context of Malaysian higher education institutions. We integrate the teachers' digital competency construct with the individual characteristic, a TTF construct, to validate our proposed framework linking with the TTF model. With the empirical validation of this integrated framework, our study will help to obtain deep insights into the Moodle utilization and performance impact from the teachers' digital proficiency, LMS, and academic tasks perspectives. It will also contribute to the digital competency research constituent by carrying out interesting implications both theoretically and practically for practitioners, academics, and policymakers.

**2. Literature Review**

*2.1. Task-Technology Fit (TTF) Model*
TTF model developed by Goodhue and Thompson [5], in a broad context of Information System (IS), explains the components of IS support users in performing their desirable tasks. The model also clarifies that users' interactions towards IS should be interlinked with their abilities to use technology. Such components consist of task characteristics, technology characteristics, and individual characteristics. The individual characteristics, elucidated by Goodhue and Thompson [5], are highlighted as the degree of human aptitudes, including but not limited to users' motivation, training needs, and IS proficiency.

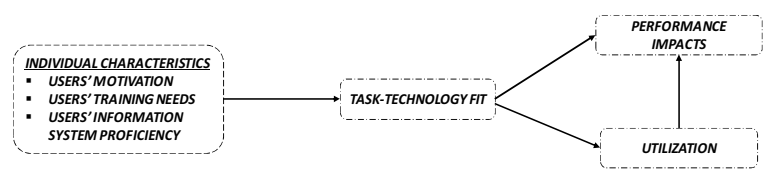

**Figure 1.** Goodhue and Thompson's TTF Model

According to the TTF model (Figure 1), individual characteristics lead to Task-Technology Fit (TTF), which in turn affects the technology utilization and individuals' performance as well. In this research, IS proficiency (digital competency), an aspect of individual characteristics, is defined as ones' technical skills and abilities, which helps to carry out a good fit of technology for systems utilizing and improving users' performance.

*2.2. Digital Competency (Individual Characteristics)*
The TTF model identifies the users' competencies for assessing and understanding technological capabilities that may support to complete their work roles. The model suggests that if technology embraces users' skills and abilities that help to fit it with their task requirements and therefore ensure their better performance as well [7]. Understanding individual characteristics in the context of technology utilization, digital competency (DC) is meant as the set of individuals' skills, abilities, styles, and overall technical knowledge, towards information technology (IT) for carrying out work roles, solving difficulties, improving, communicating, and managing information, collaborating, forming and sharing documents, and developing knowledge, which involves ones' values, responsiveness, interaction, innovativeness to learn, work, and participate in the digital environment [13], [15]. According to Omotayo and Haliru [7], technology proficiency (digital competency) has been a vital and well-accepted phenomenon in IS research. It is a broader concept, as an individual characteristic, used in the TTF theory that has been used pertinent in various IS study context. In our study, DC contains four particular components, namely technology literacy, knowledge deepening, presentation skills, and professional skills.

The basic technology skill, such as being able to use and manage any technology or tool, is referred to as technology literacy (TL) [13]. The TTF model theorizes that when users have enough abilities to use technology, it will be fit for the tasks it supports [5]. This means that if users' competencies meet technological functionality, the technology will fit to complete their tasks [10], [16]. Mohammadyari and Singh [17] suggested that users' digital literacy (technology literacy) enables using learning materials that should be considered when assessing their influence on task-performance.

Knowledge deepening (KD), a key component related to teachers' IS proficiency, refers to the teachers' mastery to correspond, understand, handle, and perform ICT activities, such as problem-solving, critical thinking, analyzing data, applying policies and skills, and managing information [13]. A survey of 186 university educators by a particular group of scholars [11] confirmed that teachers' mastery of ICT knowledge positively and significantly influenced their intention using ICT ($\beta = 0.37$, $p < 0.01$), and this intention had a direct and significant effect on ICT utilization ($\beta = 0.13$, $p < 0.01$). Another study with a sample size of 95 academic staff proved that when users' are confident about their knowledge using technology, they are more willing to use technology for sharing their knowledge, resulting in higher TTF variation (37%) [4].

On another note, teachers' didactic instructions, known as presentation skills (PRS), is the way of teaching skills, such as explaining to students regarding the topic; discussing, elucidating, representing, lecturing, posing inquiries to them, responding to their queries, and conversing with them [13]. According to Curran-Everet [18], users' better presentation skills improve a better ability to engage and connect with their activities, whatever the situation it is. Based on the mixed-method approach, a more recent study showed that users' presentation skills could overcome the difficulties of technology usage, which, in turn, positively affected their task performance [6].

Another dimension of DC, which teaching staffs require in their work is professional skills (PFS). The teachers' professional skills are their acquired knowledge and experience in different ways, such as any course, program, conference, seminar, event and workshop, experimentation, observation, professional network, and relation [13]. Conducting a semi-structured interview, Akkuzu [19] found that the interviewers, who were pleased with their prior ICT skills, or who were comfortable with the previous technology, are more skillful in using any new material than a novice user. According to a group of scholars [20], acquired experiences are directly associated with innovative tasks. However, academics proved that previous technology experience has a robust correlation with human-technology interaction, which meets the task requirements [21].
With this understanding, the following hypotheses are thus proposed:

Hypothesis (H1): Technology Literacy in using Moodle has a positive effect on teachers' TTF.
Hypothesis (H2): Knowledge deepening towards using Moodle positively affects teachers' TTF.
Hypothesis (H3): Presentation skills for using Moodle positively influences Teachers' TTF.
Hypothesis (H4): Professional skills in using Moodle has a positive impact on Teachers' TTF.

*2.3. Task-Technology Fit (TTF)*
Technological abilities performing individuals' tasks are described by the well-accepted model named Task-technology fit (TTF), explained as the extent to which the capability of technology should fit the individuals' needs to support their tasks [5]. In more detail, TTF is the ability of technology that links between user capabilities, task demands, and technological functions as well [10]. According to Diar, Sandhyaduhita, and Budi [22], studying to understand human interactions towards using information systems (IS) is paramount. Hence, in emerging the users' approach towards IS technology, this study is guided by the TTF model in the existing literature to identify TTF as an integrally consistent forecaster of educators' Moodle usage and their task performance after using it.

　　The TTF Model underlines that IS must match the tasks it performs, which positively impacts users' performance [5]. The study proved that if the technology fits the task at hand, it will support executing individuals' tasks more simply and efficiently, which in turn, further will increase their task performance [23]. Additionally, TTF is associated with individuals' usage of IS, which is essential in a broad context in a viewing of impacting performance criteria [9]. Within this context, prior researchers empirically evidenced on using different IS technology for conceptualizing a good fit between IS functionality and its usage to execute an individual's tasks. For example, Tam and Oliveira [24] proved when the technology fits the task requirements, it positively affects users' interaction using technology, resulting in their better performance. Hence, it is highly plausible that TTF is a positive and significant predictor of both systems' utilization and users' performance[7], [4], [5], [8], [22], [25]. From this perspective, we hypothesize the subsequent tentative statements:
Hypothesis (H5): TTF positively influences Moodle's utilization by teachers.
Hypothesis (H6): Teachers' performance by using Moodle is positively affected by TTF.

*2.4. Utilization (Use of Moodle)*
The TTF theory defines utilization as users' behavior of using information systems (IS) [5]. The model further explains that TTF must be users' determinate of positive views on the effectiveness and value of the systems and the tasks well-performed by using technology. In our study, utilization is referred to as the Moodle using by teaching staff at the university level to complete their tasks. As Moodle, considering an effective VLE, is very helpful for academics, its utilization must be effective. When Moodle will be a good fit between teachers' proficiencies and their usage, their performance will be much better. However, the prior researchers evidenced that users' utilization with higher technology fit positively affected performance impacts [4], [5], [8], [22], [24]. Within this context, the next hypothesis has been developed.
Hypothesis (H7): Utilization of Moodle positively influences teachers' performance.

**3. Research Framework**
The research framework for this study has drawn on insights from Goodhue and Thompson's TTF theory [5]. Digital competency in our context compresses as a constituent dimension of individual characteristics, one of the subsets of the TTF model. Our framework discourses that digital competency has deeply rooted in four key points, namely, technology literacy, knowledge deepening, presentation skills, and professional skills. Within this context, our study framework (Figure 2) specifies that such digital competencies affect teachers' task-technology fit, which in return would impact both system's utilization (use of Moodle) and performance impacts (teachers' performance), whereas utilization forecasts performance impacts.

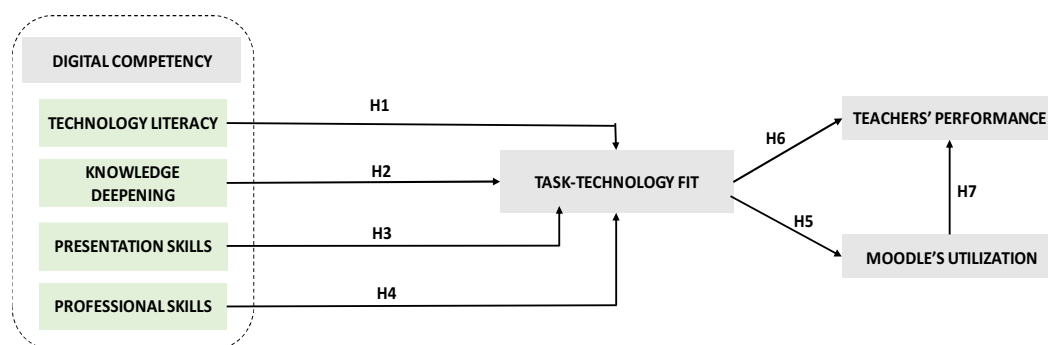

**Figure 2.** Proposed Research Framework

### 4. Methodology

A quantitative research method has been adopted, which deals with the positivist paradigm, deductive approach, cross-sectional approach, and survey strategy [26]. Based on the positivist paradigm, this study seeks to link causal relationships that how independent variables cause a change in their dependent variables. We applied Goodhue and Thompson's TTF theory [5] for testing the proposed hypotheses, which has been labeled as a deductive approach. On the other hand, as this study involved collecting data at one point in time, that's why it is integrally cross-sectional in nature.

Administered to mainly academic professionals including teaching staff at the university level in Klang Valley, an online survey was derived using purposive sampling. During this current pandemic (COVID-19), it was quite tough to collect the data from respondents. Hence, purposive sampling was the most time-effective sampling technique for this study to identify conceivable participants [27]. To reach the respondents, we communicated with the key contact persons of several universities by conferring the study's aims and scopes. They were convinced to invite all participants via an online platform to fill out the survey questionnaire. The survey lasted over eight weeks. In the end, a total of 238 responses completed the survey anonymously.

The Questionnaire section was separated into two categories. The former focused on defining demographic variables, whereas the latter one was involved with items related to the main study's variables. Demographic variables consist of gender, age, educational background, and teaching area. To measure most constructs in this research, the items were adapted, which were previously deep-rooted in the existing studies. The three key components of digital competency were assessed with 9 items adapted from existing literature, including technology literacy (3 items) [28], [29]; knowledge deepening (3 items) [29]; and presentation skills (3 items) [6]. For developing scales for professional skills, this study pursued established guidelines in the prior researcher's work [19]. On the other side, five items, three items, and four items were adapted for measuring TTF [5], utilization (use of Moodle) [8], and teachers' performance [5], respectively. Each item was measured using a five-point Likert Scale, anchored from 1 (strongly disagree) to 5 (strongly agree).

### 5. Data Analysis Procedures and Results

The data analysis tools, namely SPSS 25.0 and AMOS 23.0 were used for analyzing the proposed hypotheses. As the analysis of a moment structures (AMOS) is an appropriate approach to test exploratory models, including manifold-valued variables with multiple items, this approach was preferred as this study's data analysis technique. It is also called causal modelling software used for determining overall Structural Equation Modelling (SEM) and confirmatory factor analysis (CFA) analysis. A two-stage approach was used for this study's model testing [30]. The former one included the measurement model assessing confirmatory factor analysis (CFA) to test the reliability and validity of each construct, and the latter one involved with the structural model to examine each proposed hypothesis.

The questionnaire data provided both respondents' demographic and statistical analysis data. In our online survey, respondents' classification showed the higher representation of male (64%) than female (36%) with a majority of 30-39 years age group (50%), while others categorized into three specific age groups, namely 20-29 years (23%), 40-49 years (18%), and above 50 years (9%). Master's degrees were obtained by most of the participants (71%), while the rest of them held the doctorate or others degree (29%). The business department had more depiction (37%) than the other respective departments, such as IT, Engineering, Medicine, and others together constitute 63%.

To build the causal relationship, structural equation modeling (SEM) is used in this study. SEM analysis is based on two models [31], [32] measurement model that indicates the items reliability, validity, and loadings towards the variables, and the structural model defines the causal relationship in the model [33]. Prior to the structural model, data needs to be valid and reliable for the process. To do so the reliability (i.e., Cronbach's Alpha >0.70) and validity analysis (Average Variance Extracted or AVE >0.50) of respondents' data were tested. As shown in Table 3, the reliability and validity of all items are fulfilling the minimum requirement as Cronbach Alpha >0.70, Composite Reliability (CR) >0.70, and AVE >0.50 [34]. The next procedure is to assess the correlation between the model variables, here in Table 1, the correlation values among the variables are less than 0.90 that ensures the unidentical in the model.

Table 1. Reliability, Validity, and Correlation Results

| | Cronbach's Alpha | CR | AVE | KD | PFS | PRS | TL | TP | TTF | UT |
|---|---|---|---|---|---|---|---|---|---|---|
| **Knowledge Deepening (KD)** | 0.77 | 0.87 | 0.68 | 1 | | | | | | |
| **Professional Skills (PFS)** | 0.76 | 0.86 | 0.67 | 0.39 | 1 | | | | | |
| **Presentation Skills (PRS)** | 0.81 | 0.89 | 0.72 | 0.53 | 0.29 | 1 | | | | |
| **Technology Literacy (TL)** | 0.82 | 0.89 | 0.73 | 0.80 | 0.36 | 0.51 | 1 | | | |
| **Teacher Performance (TP)** | 0.81 | 0.88 | 0.64 | 0.48 | 0.45 | 0.35 | 0.52 | 1 | | |
| **Task-Technology Fit (TTF)** | 0.89 | 0.92 | 0.70 | 0.80 | 0.50 | 0.51 | 0.75 | 0.64 | 1 | |
| **Utilization (UT)** | 0.82 | 0.89 | 0.74 | 0.34 | 0.50 | 0.24 | 0.35 | 0.48 | 0.45 | 1 |

To proceed with the causal analysis, outer loadings of items towards respective variables were analyzed against the minimum value of 0.60 and resulted Table 2 showed the values are precise for the causal analysis[33]. In the final step prior to regression analysis of studied factors, the model fit indices are to be tested. In our analysis, all the recommended parameters [33] of model fit indices were achieved as exhibited in Table 3.

In the structural model, causation was tested by regression analysis, and hypotheses testing was assessed by the t-statistics threshold i.e., t >1.96, and level of significance i.e., p-value < 0.05 [27]. As shown in Table 4, all the hypotheses proved significant towards perspective paths with t-statistics higher than 1.96 and p-value less than 0.05 and contain the significant regression weights i.e., more than 0.20.

Table 2. Outer Loadings

| Variable | Item | Outer Loadings | Variable | Item | Outer Loadings |
|---|---|---|---|---|---|
| **Technology Literacy (TL)** | TL1 | 0.852 | **Teachers Performance (TP)** | TP1 | 0.817 |
| | TL2 | 0.855 | | TP2 | 0.807 |
| | TL3 | 0.862 | | TP3 | 0.795 |
| **Knowledge Deepening (KD)** | KD1 | 0.848 | | TP4 | 0.786 |
| | KD2 | 0.863 | **Task-Technology Fit (TTF)** | TTF1 | 0.799 |
| | KD3 | 0.766 | | TTF2 | 0.833 |
| **Presentation Skills (PRS)** | PRS1 | 0.826 | | TTF3 | 0.791 |
| | PRS2 | 0.848 | | TTF4 | 0.902 |
| | PRS3 | 0.877 | | TTF5 | 0.850 |
| **Professional Skills (PFS)** | PFS1 | 0.800 | **Utilization (UT)** | UT1 | 0.881 |
| | PFS2 | 0.860 | | UT2 | 0.859 |
| | PFS3 | 0.796 | | UT3 | 0.835 |

Table 3. Model Fit Indices

| Measure | Estimate | Threshold | Interpretation |
|---|---|---|---|
| CMIN | 460.695 | -- | -- |
| DF | 239 | -- | -- |
| CMIN/DF | 1.928 | Between 1 and 3 | Excellent |
| CFI | 0.924 | >0.95 | Acceptable |
| SRMR | 0.080 | <0.08 | Acceptable |
| RMSEA | 0.063 | <0.06 | Acceptable |

Table 4. Hypothesis Testing

| | Hypotheses | Standardized Regression Weights | R Square | T-Statistics | P-Value | Remarks |
|---|---|---|---|---|---|---|
| H1 | TL→ TTF | 0.25 | | 2.288 | 0.013 | Accepted |
| H2 | KD→ TTF | 0.49 | 0.713 | 4.745 | 0.000 | Accepted |
| H3 | PRS→ TTF | 0.27 | | 2.908 | 0.009 | Accepted |
| H4 | PFS→ TTF | 0.21 | | 2.693 | 0.011 | Accepted |
| H5 | TTF→ TP | 0.64 | 0.451 | 4.395 | 0.000 | Accepted |
| H7 | UT→ TP | 0.24 | | 2.176 | 0.018 | Accepted |
| H6 | TTF→ UT | 0.45 | 0.200 | 3.775 | 0.000 | Accepted |

## 6. Discussion

This study was aimed to comprehend the individual characteristics in the digital skills competency context for a better fit of technology for teaching purposes. The data analysis explored the reliable, valid, and fit data parameters of 238 questionnaires. The SEM results portrayed the better fit of task and technology from digital competency factors towards VLE usability. The total impact due to digital competency factors i.e., technology literacy (TL), knowledge deepening (KD), presentation skills (PRS), and professional skills (PFS) resulted in a 70% change in task-technology fit (TTF). This combination of individual characteristics in the TTF model is the novel contribution of this research where new dimensions of digital competency are analyzed through computer efficiency skills i.e., TL, using these skills to make the learning environment i.e. KD, then presenting the information through technology tools that trigger the learning stimuli among the students i.e., PRS and finally keeping the institution policies and strategies to make better use of technology i.e., PFS. Among these competency factors, the knowledge deepening (KD) variable had a higher regression impact (i.e., 0.49) through H2 on TTF, it shows the educators whose ability to establish the student-centered classrooms and involve the learners collaboratively, are more inclined to use the Moodle in productive performance scenario. The validity of this study's digital competency factors towards TTF signals the attentiveness policies of institutions to not merely equip the educators with IT proficiency but to guide them for making better use of technology to become digital dexterous educators [2]. The impact of TTF on Moodle Utilization is significant with 45% regression weight and 20% R-square change while the overall impact of TTF and Utilization on educator's performance (TP) is significant. The model R-square is calculated as 45.1% that means all the factors of the TTF model have a positive and healthy effect on the performance of the teacher in Moodle usability.

The study was conducted in the Malaysian context, enabling the teacher's competency in digital learning and development will ensure the virtual learning environment (VLE). The role of the educator is quite significant to mitigate the impact of virtual classrooms on learners in terms of lack of attention, screen boredom with non-interactive sessions, technical glitches, etc. In terms of presentation skills in VLE the interactive online classes can be targeted, and students' interest can be vitalized with a higher level of digital competency among educators. Higher education institutions have turned their attention to educators' digital competency for advancing their teaching-learning process from traditional approaches. Though many educators use Moodle, an open-source VLE, as a repository for instructional materials, they do not understand the different functionalities that it offers. Educators' technology proficiency will help them to be involved in the teaching activities pedagogically, enable them to be innovative to use Moodle, and more simply and swiftly complete their tasks and efforts efficiently [35]. With the contributions of our research, academics and policymakers will comprehend thus innovative ideas, knowledge, and skills for Moodle usage in the teaching process.

## 7. Conclusion

In the information age, there are different witnessed trends that information systems (IS) are inevitable in the current teaching system. A Virtual Learning Environment (VLE) such as Moodle has become a vital part of the teaching practices. Given this aspect, this article aimed to examine the educators' digital competencies of using Moodle and a good fit of Moodle (TTF) towards their Moodle's utilization and their task performance in Malaysian higher education institutions context. In SEM, our analyses revealed that all the proposed constructs robustly significantly positively associated with its respective relationship. These findings can provide a better understanding of educators' technology proficiency towards Moodle usage, which may lead to the advanced grounding to overall strategic planning and policy in their activities. Furthermore, our results can corroborate the TTF theory in defining DC as individuals' capability using any technology.

## 8. Research Limitations and Further Recommendations

This study includes several common limitations and further recommendations. Firstly, the concept of DC has been generalized based on UNESCO's ICT Competency Framework for Teachers [13]. However, the framework highlights 18 proficiencies based on six features of using IS in educators' professional practice. Based on this, we integrate only teachers' overall technology proficiency with digital competency while using VLE like Moodle. But knowledge management, a crucial aspect of digital competency, enables educators to facilitate any learning environment [11], [13]. Hence, further studies can focus on this particular phenomenon in the context of embracing virtual education systems and pondering over the usage enabling features such as system and service quality of VLE can also yield useful insight [36]. On the other hand, we have reconciled the digital competency construct, an individual characteristic, with the TTF model underlying a solid theoretical foundation. On the contrary, individual characteristics associated with the TTF-related construct also encompass users' motivation and training needs[5], [35]. Individual-oriented approaches, integrating both factors can be addressed more comprehensively in any further research. A non-representative sample (purposive sampling) may be an unreliable source to draw the study population, which can lead to a possible sampling bias [37]. Hence, any representative sampling technique (i.e., stratified random sampling, systematic random sampling) should be preferred by future researchers. Lastly, though the qualitative approach is used in previous studies' research design regarding research in VLE in the academic context, these are quite rare examples. Therefore, further studies should tend for more qualitative research methods.